\documentclass[pra,aps,twocolumn,superscriptaddress,nofootinbib]{revtex4-1}

\usepackage{amsmath}  \usepackage{amssymb}  \usepackage{amsfonts}  \usepackage{bm}  \usepackage{bbm}   \usepackage{braket}   \usepackage{comment}  \usepackage{dcolumn}  \usepackage{enumerate}  \usepackage{epsfig}  \usepackage{gensymb}  \usepackage{graphicx}  \usepackage{indentfirst}  \usepackage{mathtools}  \usepackage{psfrag}  \usepackage{pst-all}  \usepackage{soul}  \usepackage{siunitx}   \usepackage{xcolor}  \usepackage{microtype} \usepackage{txfonts} \usepackage{txfontsb} 
\usepackage{float} 
\usepackage[colorlinks,linkcolor=blue,citecolor=blue,urlcolor=blue]{hyperref} 
\usepackage[T1]{fontenc} 

\def\ii{{\rm i}}  \def\ee{{\rm e}}
\def\me{m_{\rm e}}  
  
\def\Ab{{\bf A}}        \def\Eb{{\bf E}}                                    \def\Rb{{\bf R}}  \def\rb{{\bf r}}      \def\vb{{\bf v}} 
\def\xx{\hat{\bf x}}    \def\zz{\hat{\bf z}}    \def\rr{\hat{\bf r}}        

\def\vv{\hat{\bf v}}    \def\db{{\bf d}}  
\usepackage{lipsum}

\begin{document}
\def\bibsection{\section*{\refname}}

\title{Rydberg-atom manipulation through strong interaction with free electrons}

\author{Adamantios~P.~Synanidis}
\affiliation{ICFO--Institut de Ciencies Fotoniques, The Barcelona Institute of Science and Technology, 08860 Castelldefels (Barcelona), Spain}
\author{P.~A.~D.~Gon\c{c}alves}
\affiliation{ICFO--Institut de Ciencies Fotoniques, The Barcelona Institute of Science and Technology, 08860 Castelldefels (Barcelona), Spain}
\author{F.~Javier~Garc\'{\i}a~de~Abajo}
\email[E-mail: ]{javier.garciadeabajo@nanophotonics.es}
\affiliation{ICFO--Institut de Ciencies Fotoniques, The Barcelona Institute of Science and Technology, 08860 Castelldefels (Barcelona), Spain}
\affiliation{ICREA--Instituci\'o Catalana de Recerca i Estudis Avan\c{c}ats, Passeig Llu\'{\i}s Companys 23, 08010 Barcelona, Spain}

\date{\today}

\begin{abstract}{\bf \noindent
Optically trapped Rydberg atoms are a suitable platform to explore quantum many-body physics mediated by long-range atom--atom interactions that can be engineered through externally applied light fields. However, this approach is limited to dipole-allowed transitions and a spatial resolution of the order of the optical wavelength. Here, we theoretically investigate the interaction between free electrons and individual Rydberg atoms as an approach to induce nondipolar transitions with subnanometer spatial precision and a substantial degree of control over the final atomic states. We observe unity-order excitation probabilities produced by a single electron for suitably chosen combinations of electron energies and electron-beam distance to the atom. We further discuss electron--atom entanglement in combination with lateral shaping of the electron followed by postselection. Our results support free electrons as powerful tools to manipulate Rydberg atoms in previously inaccessible ways.
}\end{abstract}

\maketitle

\section{Introduction}

Rydberg atoms have an electron in an orbital with a high principal quantum number~\cite{G94} and display unique properties that find application in quantum computing~\cite{SWM10,GKG19,LKS19,APS20} and quantum simulation of many-body systems~\cite{BL20,LBR16,BSK17,WML10,DLS19,SSW21,EWL21}. In particular, highly excited Rydberg atoms possess long lifetimes ($\lesssim1\,$ms) and large dipole moments, rendering them highly sensitive to external fields and enabling long-range atom--atom interactions~\cite{SW05_2,BRT09,HGP24}. Because of the latter, the excitation of an atom to a Rydberg state inhibits the subsequent excitation of nearby atoms~\cite{LFC01,UJH09,GMW09}, thus enabling the implementation of quantum gates and entanglement~\cite{IUZ10,WGE10}, while also producing a strong nonlinear optical response at the few- and single-photon level~\cite{GOF11,PFL12,FPL13}.

The formation and study of quantum gases of ultracold Rydberg atoms trapped in optical lattices~\cite{BDL16,BL20} are currently relying on optical techniques, which are employed to coherently control atom--atom interactions~\cite{SCE12,LBR16} as well as to image~\cite{BGP09,SWE10_2,NLW07} and read out the system~\cite{NLW07,SWM10}. However, far-field excitation and probing restrict the control schemes to dipole transitions with a spatial resolution limited by optical diffraction.

Electron beams (e-beams) offer an appealing alternative for targeting and manipulating atomic and molecular transitions~\cite{MN1934,RE1965,THB21,paper393}, where the evanescent electromagnetic field associated with free electrons mediates the interaction~\cite{paper149}. This field is highly inhomogeneous near the e-beam and contains multipolar components capable of inducing nondipolar transitions inaccessible to far-field light. In addition, e-beams can be focused down to subnanometer lateral widths~\cite{paper149,paper371}, thus providing an enhanced spatial resolution to address individual atoms. In the context of quantum gases, broad e-beams have been used as probes to image atoms in optical lattices~\cite{GWR08,WLG09,MWN14}, while quasi-free Rydberg electrons have been demonstrated to couple to Bose--Einstein condensates~\cite{BKG13}. The use of e-beams to control and study Rydberg atoms could further benefit from recent advances in modulating the free-electron wave function through the interaction with free-space~\cite{paper368,MWS22} and scattered~\cite{BFZ09,FES15,PRY17,RK20,YFR21,paper315,paper351,paper397,paper437} optical fields or via electrostatic electron phase shaping~\cite{VBM18,YVB23}.

\begin{figure*}[ht!] 
\centering\includegraphics[width=1.0\textwidth]{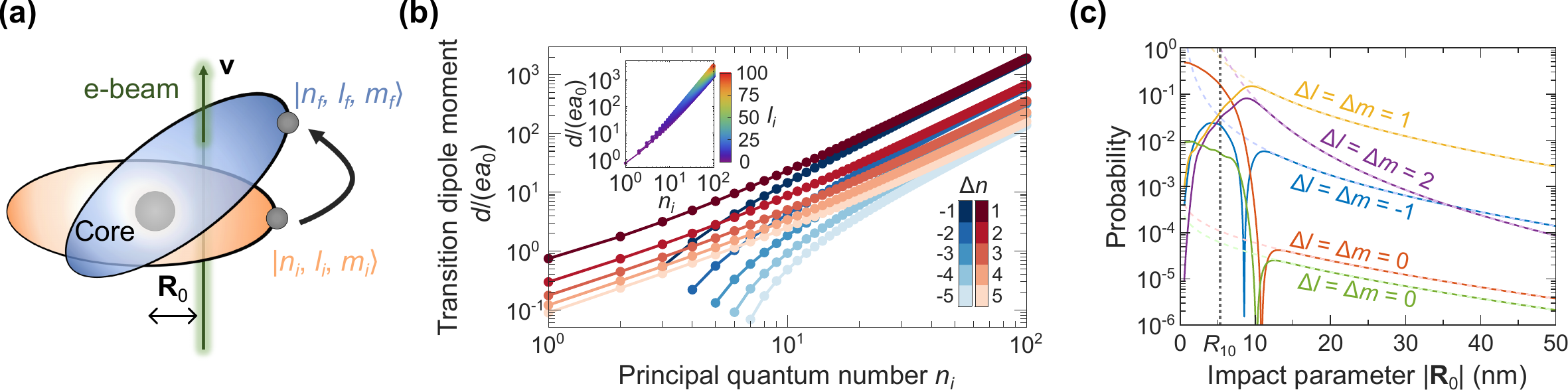}
\caption{\textbf{Interaction of Rydberg atoms with electron beams.}
\textbf{(a)}~The interaction of a Rydberg atom with a free electron moving along a straight line and passing at a position $\Rb_0$ from the nucleus (impact parameter) can drive transitions between initial and final states, $\ket{n_i,l_i,m_i}$ and $\ket{n_f,l_f,m_f}$, respectively.
\textbf{(b)}~Transition dipole moment $d=e\big|\langle n_i,0,0|\rb|n_f,1,1\rangle\big|$ (in units of $ea_0$) associated with dipolar transitions in the Rydberg atom as a function of initial principal quantum number $n_i$ for different values of $\Delta n=n_f-n_i$ (see color legend). The inset shows results for $\ket{n_i,l_i,0} \to \ket{n_i+1,l_i+1,1}$ transitions with a wide range of $l_i$ values.
\textbf{(c)}~Impact-parameter dependence of the probability of $\ket{10,5,5} \to \ket{11,5+\Delta l,5+\Delta m}$ transitions for different values of $\Delta l$ and $\Delta m$ as produced by interaction with a 10~eV electron. First-order matrix-element calculations (solid curves) are compared to the large-$|\Rb_0|$ scaling $\propto K^2_{|\Delta m|}(|\varepsilon_{fi}|R_0/v)$ with $\hbar\varepsilon_{fi}=(1/n_i^2-1/n_f^2)$~Ry (dashed curves). A vertical broken line indicates the atomic radius $R_n=n^2a_0$ with $n=10$ and $a_0$ the Bohr radius.}
\label{Fig1}
\end{figure*}

Here, we introduce a new approach for manipulating Rydberg atoms using free electrons (Fig.~\ref{Fig1}a). By characterizing the e-beam--Rydberg-atom interaction theoretically, we reveal a wealth of nondipolar transitions reaching the strong-coupling regime (i.e., near-unity transition probabilities). We demonstrate that the final state of the Rydberg atom consists of energy and angular-momentum superpositions that are strongly dependent on the e-beam energy and impact parameter (distance to the atom), thus enabling the coherent control of individual atoms. Through interaction with a periodic train of free electrons, precise transition frequencies are selected, corresponding to the delay between consecutive electrons. A wide range of final atomic states, including entanglement among two or more Rydberg atoms, is reached by using laterally shaped electron waves followed by electron postselection. Our results highlight appealing capabilities of e-beams for manipulating quantum systems based on Rydberg atoms in unprecedented ways using currently available state-of-the-art techniques.

\section{Results}

\subsection{Theoretical framework}

We describe the electron as a classical point particle following a straight-line trajectory with a constant velocity vector $\vb$ (nonrecoil approximation~\cite{paper149}), as illustrated in Fig.~\ref{Fig1}a. This approach is justified under the assumption that the e-beam is focused laterally down to a small diameter (e.g., $<1$~nm in transmission electron miscroscopes) compared to the size of the Rydberg atom (e.g., $\sim5-500$~nm for principal quantum numbers $n=5-50$). In addition, the excitation probability is known to be independent of the electron wave function along the e-beam direction~\cite{paper371}, which justifies treating the electron as a classical point particle to make theory simple. The moving electron introduces an external electromagnetic field that acts on the Rydberg atom, and we study their quantum-mechanical interaction by adopting the minimal coupling prescription~\cite{S94} and neglecting $A^2$ terms. Then, the Hamiltonian of the system reduces to $\hat{\mathcal{H}}(\rb,t)=\hat{\mathcal{H}}_0(\rb)+\hat{\mathcal{H}}_1(\rb,t)$, where $\hat{\mathcal{H}}_0(\rb)= -(\hbar^2/2\me)\nabla^2+V(\rb)$ describes the Rydberg electron in a Coulomb-like core potential $V(\rb)$ while $\hat{\mathcal{H}}_1(\rb,t)=-(\ii e \hbar/2\me c) \big[ \bm{\nabla} \cdot \Ab(\rb,t) + \Ab(\rb,t) \cdot \bm{\nabla}\big] - e\phi(\rb,t)$ accounts for the perturbation produced by the swift electron through the vector and scalar potentials $\Ab$ and $\phi$. Writing the electron trajectory as $\rb_0(t) = \Rb_0 + \vb t$ (with $\Rb_0\perp\vb$) and its associated electric field as $\Eb(\rb,t) = -(1/c)\, \partial_t\Ab(\rb,t) - \bm{\nabla}\phi(\rb,t)$,
we have the potentials~\cite{J1975}
\begin{align}
\phi(\rb,t) &= - \frac{e \gamma}{\sqrt{|\Rb - \Rb_0|^2 + \gamma^2 (\rb\cdot\vv - vt)^2}} , \nonumber
\end{align}
and $\Ab(\rb,t)=(\vb/c)\,\phi(\rb,t)$, where $\gamma = (1-v^2/c^2)^{-1/2}$ is the Lorentz factor. In what follows, we consider low-energy electrons ($v\ll c$), so we approximate $\gamma\approx1$ and neglect the effect of the vector potential.

In general, the wave functions of the Rydberg states can be calculated analogously to those of the hydrogen atom using the so-called quantum defect theory~\cite{S1983}, where the differences between the Coulomb potential of the proton in hydrogen and the actual potential experienced by a Rydberg electron are conveniently encapsulated in a species-dependent model potential with empirical parameters~\cite{G94,S1983,MSD94}. For simplicity, in this work, we consider Rydberg states described by hydrogenic wave functions $\psi_i(\rb)$ and neglect fine and hyperfine structure effects~\cite{LDC05}, so they satisfy $\hat{\mathcal{H}}_0(\rb)\, \psi_i (\rb,t)= \hbar\varepsilon_i\, \psi_i(\rb,t)$ with $V(r)=-e^2/r$ and eigenenergies $\hbar\varepsilon_i=(-1/n_i^2)\,$Ry expressed in terms of the Bohr radius $a_0$. Here, $i\equiv(n_i,l_i,m_i)$ encapsulates the principal, angular, and azimuthal quantum numbers. Hydrogenic states form a complete orthonormal basis set, and therefore, the time-dependent wave function can be expanded as $\psi(\rb,t) = \sum_i \alpha_{i}(t)\, \psi_{i}(\rb)\,\ee^{-\ii \varepsilon_i t }$, where the expansion coefficients $\alpha_{i}(t)$ satisfy the equation of motion
\begin{align}
\dot{\alpha}_i(t) = -\frac{\ii}{\hbar} \sum_{j \neq i} \ee^{\ii \varepsilon_{ij} t}\, M_{ij}(t)\, \alpha_j(t) 
\label{eq:SE-alphas}
\end{align}
with energy differences $\varepsilon_{ij} = \varepsilon_{i} - \varepsilon_{j}$ and $M_{ij}(t)$ 
and transition matrix elements defined by $M_{ij}(t) = \int \mathrm{d}^3\rb\, \psi_i^*(\rb)\, \hat{\mathcal{H}}_1(\rb,t)\, \psi_j(\rb)$ (Fig.~\ref{Fig1}a). [See Appendix~\ref{appendixA} for details on the solution of Eq.~(\ref{eq:SE-alphas}).] This formalism can be straightforwardly applied to Rydberg electron wave functions resulting from more complex phenomenological core potentials $V(r)$~\cite{G94,S1983,SPA17}.

\subsection{Interaction of a free electron with a Rydberg atom}

Before solving the full dynamics described by Eq.~\eqref{eq:SE-alphas}, it is instructive to examine the interaction within first-order perturbation theory, where the transition probability becomes
\begin{align}
P^{(1)}_{i \rightarrow f} &=  \frac{1}{\hbar^2}\bigg| \int_{-\infty}^{\infty} \mathrm{d} t\, \ee^{\ii \varepsilon_{fi} t}M_{fi}(t) \bigg|^2 \nonumber\\
&=\bigg(\frac{2e^2}{\hbar v}\bigg)^2 \bigg|\!\int_{-\infty}^{\infty} \!\!\! \mathrm{d}^3\rb\, \psi_f^*(\rb)\, \psi_i(\rb)\, \ee^{\ii\varepsilon_{fi}\rb\cdot\vb/v^2} K_0\bigg(\frac{|\varepsilon_{fi}||\mathbf{R}-\mathbf{R}_0|}{v}\bigg) \bigg|^2.
\nonumber
\end{align}
The electric dipole moment of a highly excited Rydberg atom exhibits a $\propto n^2$ scaling with the principal quantum number, reaching large values for $n \gg 1$ (Fig.~\ref{Fig1}b). In the dipolar limit (i.e., for e-beams passing far away from the atom, so that the interaction is dominated by dipolar transitions), the probability reduces to~\cite{paper149,paper371}
\begin{align}
P^{(1)}_{i \rightarrow f}=\bigg(\frac{2e \varepsilon_{fi}}{\hbar v^2}\bigg)^2 \Bigg[\big|d_{fi,x}\big|^2  K_1^2\bigg(\frac{|\varepsilon_{fi}|b}{v}\bigg)+\big|d_{fi,z}\big|^2  K_0^2\bigg(\frac{|\varepsilon_{fi}|b}{v}\bigg)\Bigg],
\label{eq:dipole_limit}
\end{align}
where $d_{fi,x}$ and $d_{fi,z}$ denote Cartesian components of the transition dipole moment $\db_{fi}=-e\int \mathrm{d}^3\rb\, \psi_f^*(\rb)\, \rb\, \psi_i(\rb)$, and we take the electron trajectory as $\rb_0(t) =  b\, \xx + v t\, \zz$. For small arguments of the modified Bessel functions $K_m$ (i.e., $|\varepsilon_{fi}|b/v\ll1$), the electron primarily couples to the $d_{fi,x}$ component (perpendicular to the e-beam), and the associated transition probabilities asymptotically approach a constant value for large principal quantum numbers $n_i$ (see Supplementary Fig.~\ref{FigS1}): even though $d_{fi,x} \propto n_i^2$, the orbital radius of the Rydberg electron also increases as $\propto n_i^2$, and thus, in the dipolar regime ($b \gg \langle r \rangle$), the increase in dipole-moment strength is compensated by the increase in impact parameter $b$. Therefore, a strong electron--Rydberg-atom interaction requires e-beams penetrating the inner region of the Rydberg atom (i.e., overlapping with the wave functions of the Rydberg electron). Indeed, Fig.~\ref{Fig1}c shows that free electrons can not only induce several types of multipolar transitions but they can also strongly couple to Rydberg atoms with near-unity probability for suitable choices of the impact parameters within the orbital radius of the Rydberg electron.

\begin{figure*}[t]
\centering\includegraphics[width=.95\textwidth]{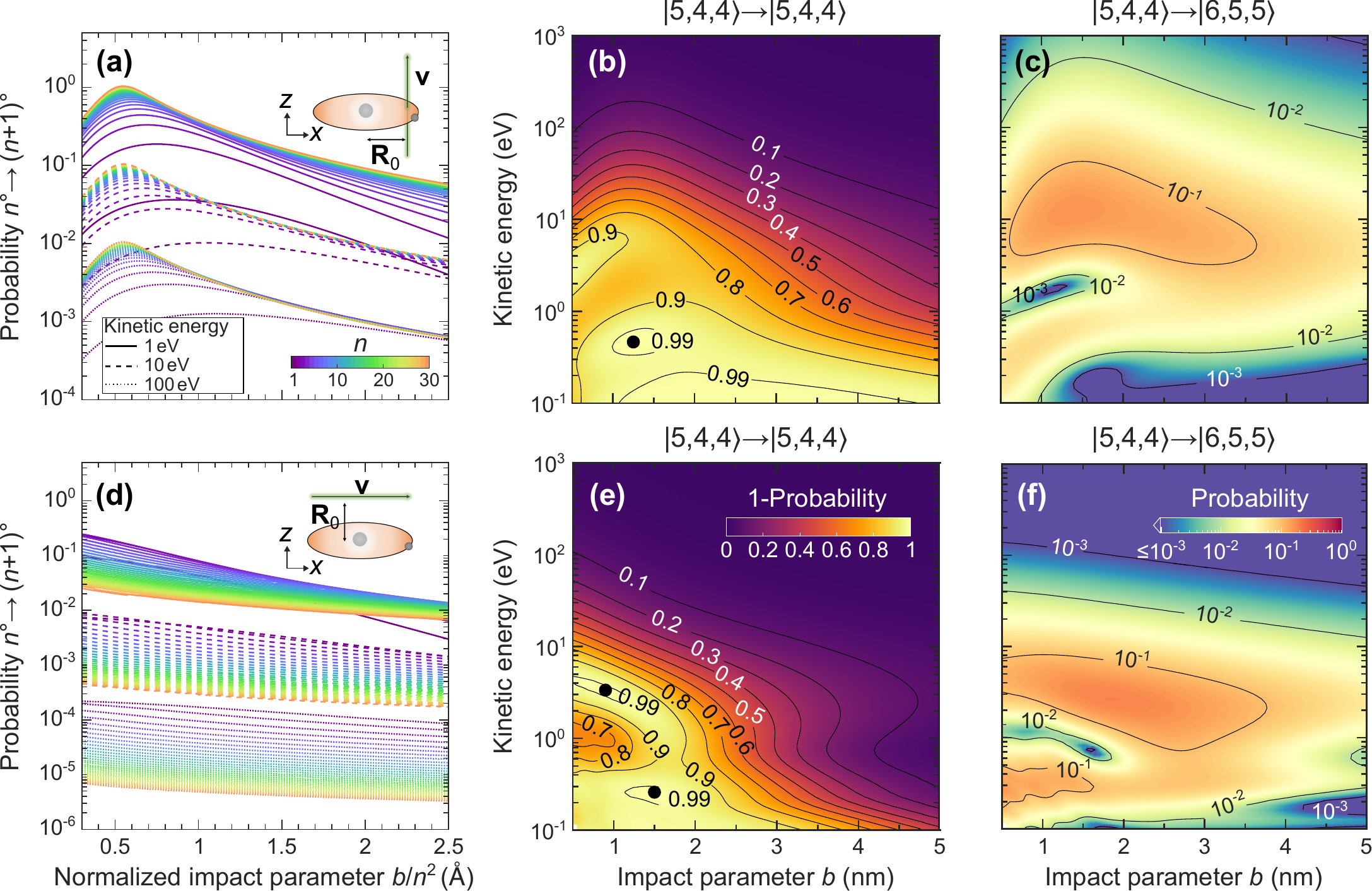}
\caption{\textbf{Strong-coupling regime.} Transition probabilities calculated for the e-beam configurations depicted in the insets of panels (a) and (d): velocity and impact parameter $\vb = v\zz$ and $\Rb_0 = b\xx$ in (a--c), and $\vb = v\xx$ and $\Rb_0 = b\zz$ in (d--f). The quantization axis is $z$ in all cases. \textbf{(a,d)}~Impact-parameter dependence of the linear-order probability for transitions $\ket{n,n-1,n-1}\rightarrow\ket{n+1,n,n}$ [i.e., between circular states $n^\circ$ and $(n+1)^\circ$] produced by free electrons of energies of $10\,$eV (solid curves), $100\,$eV (dashed curves), and $1\,$keV (dotted curves). \textbf{(b,c,e,f)}~Depletion probability of an initial state $\ket{5,4,4}$ (b,e) and probability of ending in a final state $\ket{6,5,5}$ (c,f) calculated to all orders of interaction as a function of impact parameter and electron kinetic energy. Black dots in (b,e) indicate local maxima of the depletion probability.} 
\label{Fig2}
\end{figure*}

\subsection{Strong electron--atom coupling}

The conditions for which strong electron--atom coupling is produced can be hinted by examining Eq.~(\ref{eq:dipole_limit}): because the probability scales as $1/v^2$ at large impact parameters, lowering the energy of the e-beam naturally leads to stronger coupling. Moreover, the probability diverges as $1/b^2$ near $b=0$, so we are led to explore regions close to the nucleus of the Rydberg atom to make the initial and final wave functions maximally overlap with the trajectory of the e-beam.

In Fig.~\ref{Fig2}, we illustrate some conditions for which strong coupling is attained. In particular, we study the interaction as a function of impact parameter $b$ for e-beams moving either parallel (Fig.~\ref{Fig2}a--c) or perpendicular (Fig.~\ref{Fig2}d-f) to the quantization axis (here chosen to be $z$). In Fig.~\ref{Fig2}a,d, we focus on transitions of the form $\ket{n,n-1,n-1}\rightarrow\ket{n+1,n,n}$ (i.e., involving circular states~\cite{HK83}) within first-order in the electron--atom interaction, which reveals a dramatic dependence on the orientation of the e-beam. This dependence includes the presence of a maximum of the transition probability as a function of impact parameter for $\vb$ parallel to the quantization axis, in contrast to a monotonic decrease with increasing $b$ for the perpendicular orientation. The position of the noted maximum, which scales as $b\propto n^2$, corresponds to the largest overlap of the e-beam with the involved Rydberg orbitals. In addition, increasing the principal quantum number roughly enhances the calculated transition probability in the parallel configuration (but with a weaker increase for $n\gtrsim30$). In contrast, a reduction in the coupling is observed for the perpendicular e-beam as $n$ increases.

A more quantitative measure of strong coupling is provided by the probability of depleting the initial atomic state after the passage of the electron. The transition probability calculated to first-order in the interaction already shows pronounced maxima with values well above unity (Supplementary Fig.~{\ref{FigS2}), indicating that we are entering a nonpertubative regime. When calculated to all orders of interaction (Fig.~\ref{Fig2}b,e), we observe regions of nearly 100\% depletion (see black dots in Fig.~\ref{Fig2}b,e). We note that these high values of the total transition probability indicate that the final state of the atom is orthogonal to the initial state. This observation is important for the entanglement applications discussed below.

The presence of strong initial-state depletion is accompanied by maxima of the transition probability to other states. In particular, for $\ket{5,4,4}\rightarrow\ket{6,5,5}$ transitions (Fig.~\ref{Fig2}c,f), we observe absolute maxima at $b\sim1-2$~nm, reaching values of $\approx16$\% and $\approx 19$\% (with $\approx12$~eV and $\approx3$~eV kinetic energy) for the parallel and perpendicular e-beam orientations, respectively. The impact parameter and energy position of the transition maxima depend on the final state, as shown in Supplementary Fig.~\ref{FigS3}, where we examine other elastic and inelastic transitions of dipolar and quadrupolar nature.

\subsection{Multipolar transitions enabled by swift electrons}

In contrast to the interaction with far-field light, which can only produce transitions subject to the dipolar selection rules, the electromagnetic field displayed by the moving electron contains highly nondipolar components that can induce transitions beyond those accessible with light. This effect is more dramatic when the free electron overlaps with the electronic orbitals. In addition, the electron can be regarded as a broadband source that can produce multiple transitions simultaneously, including an intrashell redistribution of the bound Rydberg electron.

\begin{figure*}[ht!]
\centering\includegraphics[width=.95\textwidth]{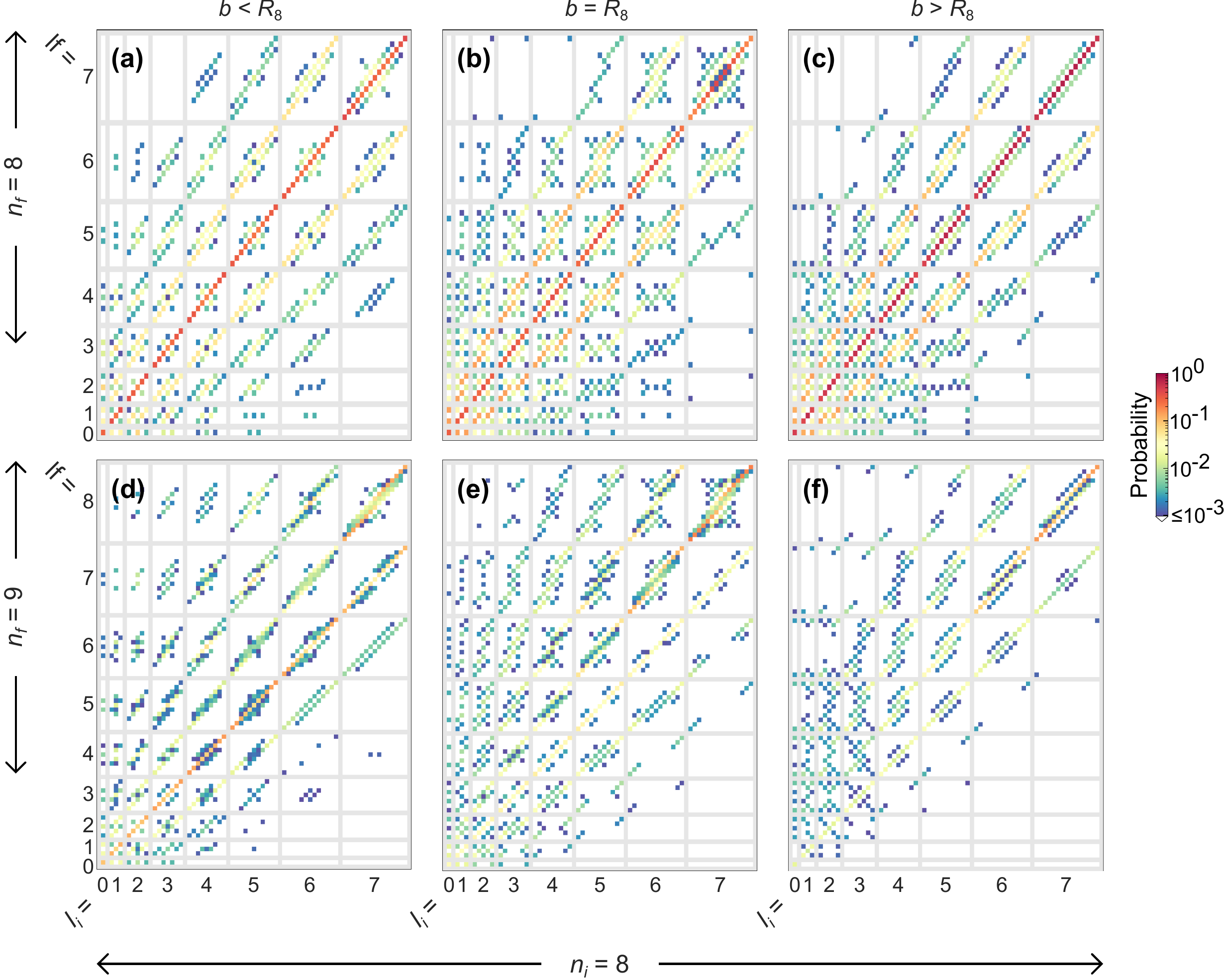}
\caption{\textbf{Multipolar excitations.} Excitation probability produced by a 10~eV electron ($\vb$ parallel to the quantization axis $z$) and calculated to all orders of interaction for impact parameters $b=R_{8}/2$, $R_{8}$, and $2R_{8}$ (left, middle, and right panels), where $R_{8}\approx3.4$~nm is the Rydberg radius for a fixed initial principal quantum number $n_i=8$. Results are shown for final states with $n_f=8$ (a--c) and $n_f=9$ (d--f). In each panel, the horizontal axis runs over initial quantum numbers $l_i$ and $m_i$, each requiring a separate simulation starting from a state $\ket{8,l_i,m_i}$. The vertical axes run over final quantum numbers $l_f$ and $m_f$. States are organized consecutively along the axes using the combined index $l^2+l+m$. Vertical and horizontal gray lines are introduced to separate rectangular areas corresponding to fixed $l_i$ and $l_f$ values.}
\label{Fig3}
\end{figure*}

Figure~\ref{Fig3} showcases the multipolar character of the Rydberg--e-beam interaction by showing the distribution of final state probabilities produced by a 10~eV electron interacting with a Rydberg atom prepared in different $n_i=8$ states. We present results calculated to all orders of scattering for electrons traversing the atom (Fig.~\ref{Fig3}a,d), passing right at the Rydberg radius (Fig.~\ref{Fig3}b,e), or well outside the atom (Fig.~\ref{Fig3}c,f). For every panel in the figure, each pixel corresponds to the transition to a given final state $\ket{n_f,l_f,m_f}$ when the atom is prepared in a state $\ket{8,l_i,m_i}$, so we organize the horizontal (initial state) and vertical (final state) axes following a normal ordering using the $l$- and $m$-dependent combined index $l^2+l+m$. For the neighboring energy transitions $n_i=8\rightarrow n_f=9$ and distances outside the electron cloud (Fig.~\ref{Fig3}f), the largest contribution comes from dipolar transitions, as we are within the range of validity of the dipolar limit. In contrast, for impact parameters at and below the Rydberg radius (Fig.~\ref{Fig3}d,e), nondipolar processes become important.

Interestingly, elastic transitions (i.e., with $n_f=n_i$) are also prominent (see Fig.~\ref{Fig3}a-c). These transitions, which cannot be accessed using a single laser source, are analogous to hybridization due to the Stark effect in the presence of a dynamical electric field (i.e., the one provided by the moving electron). Although the electron energy is preserved, we note that the electron can impart a change in angular momentum to the atom, and therefore, momentum conservation is expected to produce electron deflection, which could be exploited to reveal the presence of these types of transitions.

Incidentally, the e-beam trajectory considered in Fig.~\ref{Fig3} is parallel to the quantization axis, but symmetry could be exploited to suppress or enhance specific transitions depending on e-beam orientation.

\begin{figure*}[ht!] 
\centering\includegraphics[width=0.95\textwidth]{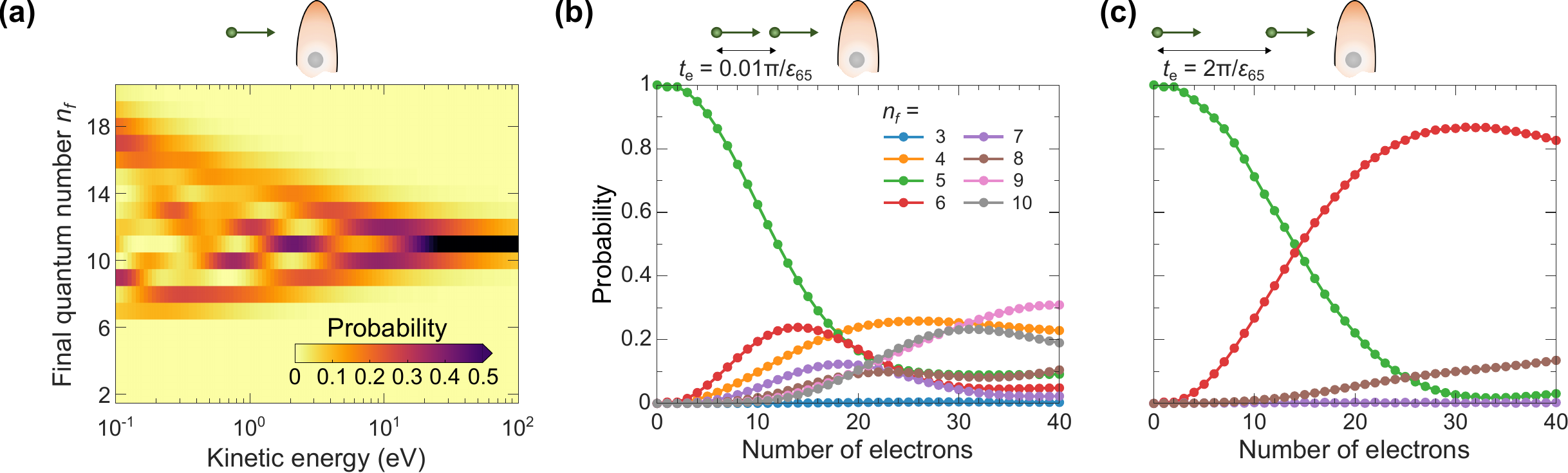}
\caption{\textbf{Tayloring the final Rydberg atom state with a periodic train of electrons.} \textbf{(a)}~Probability of finding the Rydberg atom with a final quantum number $n_f$ for an initial state $\ket{11,10,10}$ after interaction with a free electron moving parallel to the quantization axis $z$ at an impact parameter $b\approx7.01$~nm. The vertical axis runs over the incident electron energy. A broader range of $n_f$'s is observed as the electron energy is reduced. \textbf{(b,c)}~Temporal evolution of a Rydberg atom prepared in an initial state $\ket{5,4,4}$ during the interaction with an e-beam consisting of a large number $N\gg1$ of 100~eV electrons separated by a constant time difference (b) $t_\mathrm{e}=0.01\pi/\varepsilon_{65}$ and (c) $t_\mathrm{e}=2\pi/\varepsilon_{65}$ (i.e., off- and on-resonance with $n_i=5\rightarrow n_f=6$ transitions, respectively).}
\label{Fig4}
\end{figure*}

\subsection{Multiple electrons}

The interaction with a single electron produces a wide distribution of final states. This effect becomes more dramatic as we approach strong-coupling conditions by reducing the free-electron energy, as illustrated in Fig.~\ref{Fig4}a, starting in the $\ket{11,10,10}$ state and reaching the strong coupling regime for energies lower than $\sim10$~eV, where the initial state is depleted in favor of an increase in the population of adjacent energy levels. Interestingly, oscillations in the initial state population are observed at lower energies. This effect is a signature of coherent quantum dynamics. This example illustrates that we lack selectivity over the final state energy when using a single free electron.

Given the long optical periods associated with transitions in Rydberg atoms, we argue that ultrashort electron pulses (e.g., sub-ps wave packets such as those available in ultrafast electron microscopy~\cite{FES15,PLQ15}) have small durations compared with such periods and, therefore, act as point particles. By sending a periodic sequence of $N$ electrons (e.g., values of $N=1-4$ have already been demonstrated in experiment~\cite{HFD23}) separated by well-defined intervals of duration $t_{\rm e}$, the probability of a weak atomic transition can be enhanced by a factor $N^2$ relative to that of a single electron under the condition that $\varepsilon_{fi}t_{\rm e}$ is a multiple of $2\pi$. This is the so-called superradiance effect~\cite{UGK98,GIF19,GY20,paper373}, illustrated for Rydberg atoms in Figs.~\ref{Fig4}b and \ref{Fig4}c for off- and on-resonance conditions. In analogy to optical excitation by successive laser pulses, the approach explored in Fig.~\ref{Fig4}c offers a way of synthesizing highly excited Rydberg atoms by employing relatively energetic electrons (far from the strong-coupling regime), whose arrival times are engineered to successively excite the system until the desired state is reached.

\begin{figure*}[ht!] 
\centering\includegraphics[width=0.95\textwidth]{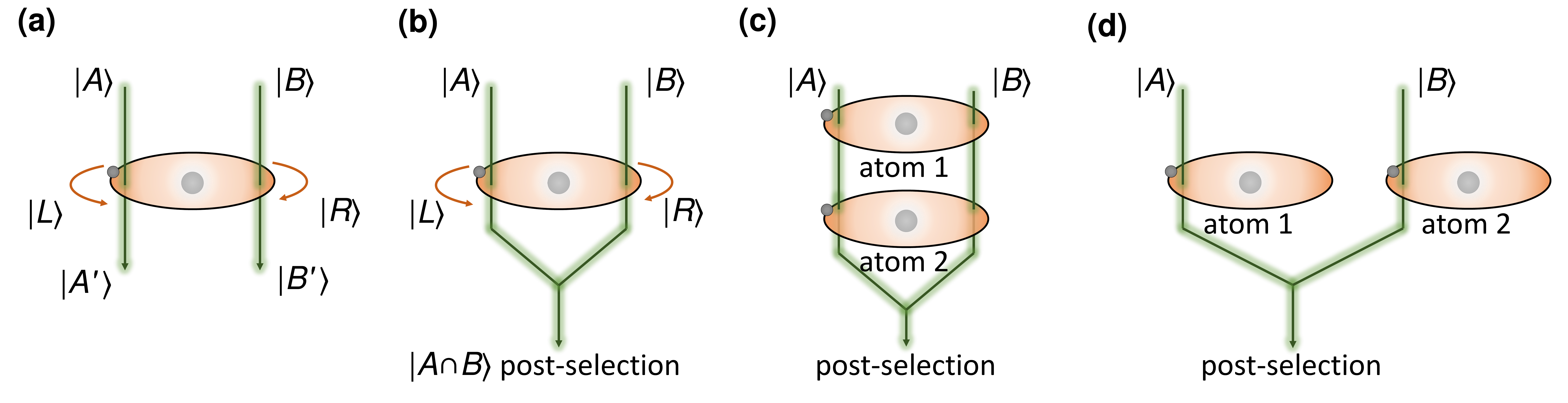}
\caption{\textbf{Quantum entanglement with free electrons and Rydberg atoms.} \textbf{(a)}~An electron prepared in a two-path state ($\ket{A}$ and $\ket{B}$) can interact with a Rydberg atom and create left and right final states ($\ket{L}$ and $\ket{R}$). After the interaction, the electron--atom system is placed in an entangled state $\ket{A'}\otimes\ket{L}+\ket{B'}\otimes\ket{R}$, where the primes indicate that the electron is also modified due to the interaction in each path. \textbf{(b)} Upon postselection of the electron considered in (a), such that one collects the contribution transmitted from both paths to a single output channel, and {\it which-way} information is erased, the atom is left in a superposition state $\ket{L}+\ket{R}$. \textbf{(c)}~After the passage of the two-path electron through two noninteracting atoms arranged in an {\it in-series} configuration, followed by electron postselection, the atomic system is left in an entangled state $\ket{L_1}\otimes\ket{L_2}+\ket{R_1}\otimes\ket{R_2}$, where the subindices refer to atoms 1 and 2. \textbf{(d)}~For two atoms in the depicted {\it in-parallel} configuration, an atomic state $\ket{L_1}\otimes\ket{I_2}+\ket{I_1}\otimes\ket{L_2}$ is obtained, where $\ket{I_j}$ denotes the initial state of atom $j$.}
\label{Fig5}
\end{figure*}

\subsection{Using free electrons to produce quantum entanglement}

Considering the large size of Rydberg atoms (radius $5.3$--$132$~nm for principal quantum numbers $10$--$50$) compared to the electron wavelength ($1.2$--$0.039$~nm at $1$--$10^3$~eV), it is physically possible to laterally focus the e-beam to form spots of small width compared to the extension of the electronic orbitals under consideration. Through the use of e-beam biprisms~\cite{MD1956,STP17} or diffraction gratings~\cite{JTM21,paper388}, the electron wave function can be made to consist of multiple paths, separated by lateral distances up to several microns. In particular, we consider a two-path e-beam arranged as depicted in Fig.~\ref{Fig5} relative to the Rydberg atom. Each of the electron paths (associated with states $\ket{A}$ and $\ket{B}$) produces a different final state of the atom (left and right, $\ket{L}$ and $\ket{R}$), so the postinteraction state of the system is entangled: $\ket{A'}\otimes\ket{L}+\ket{B'}\otimes\ket{R}$ (Fig.~\ref{Fig5}a), where the prime indicates the change produced in the electron along each path due to the interaction.

An interesting scenario is presented when {\it which-way} information is erased through electron postselection \cite{HJR24}. This can be realized by coupling each of the paths to a common single output channel. For example, if the two paths are produced by diffraction in a transmission grating, a second conjugated grating can be placed at a suitable position in which the two paths have diverged sufficiently to overlap with each other in space, such that they are both diffracted to a common Bragg beam~\cite{JTM21,paper388}. Obviously, electrons transmitted in that beam do not contain {\it which-way} information (because of the conjugation between angles of propagation of the two paths), and therefore, the detection of an electron within the angular range spanned by this beam should herald a final atomic state $\ket{L}+\ket{R}$ (Fig.~\ref{Fig5}b).

Leveraging these tools of e-beam splitting, projection, and postselection, in combination with state-of-the-art atomic optical trapping, one can produce entanglement among two or more Rydberg atoms by making the electron paths interact with them. For the {\it in-series} interaction with two atoms depicted in Fig.~\ref{Fig5}c, the $A$ path produces left states on both of the atoms, whereas the $B$ path induces right states. Upon detection of a postselected electron in the {\it which-way}-eraser configuration discussed above, the atomic system is finally prepared in an entangled state $\ket{L_1}\otimes\ket{L_2}+\ket{R_1}\otimes\ket{R_2}$, where the subindices label the two atoms.

Analogously, in an {\it in-parallel} atomic configuration (Fig.~\ref{Fig5}d) with paths $A$ and $B$ passing on the left of atoms 1 and 2, respectively, electron postselection leads to a final atomic state $\ket{L_1}\otimes\ket{I_2}+\ket{I_1}\otimes\ket{L_2}$, where $\ket{I_j}$ denotes the initial state of atom $j=1,2$. We recall that a single electron can deplete the initial state of the atom (see Fig.~\ref{Fig2}), and therefore, for a suitable combination of electron energy and impact parameter, we can have orthogonal states satisfying $\langle I_j|L_j\rangle=0$.

\section{Discussion}

We have presented an unexplored approach for manipulating the states of Rydberg atoms through their interaction with e-beams. In particular, we have theoretically shown that low-energy free electrons are capable of producing unity-order changes in the state of a Rydberg electron, including a complete depletion of the initial atomic state after interaction with a single free electron, as well as a large contribution of optically forbidden nondipolar transitions on par with (or even exceeding) dipolar ones. Because the de Broglie wavelength of the free electron is small compared with the size of Rydberg orbitals for the kinetic energies under consideration, the e-beam can be regarded as a line-like dipolar excitation oriented along the electron velocity direction~\cite{paper149}. In practice, currently avaiplable electron optics methods enable the focusing of few-eV electrons down to few-nm lateral sizes (e.g., in low-energy electron microscopes~\cite{B94,R95}). In addition, for high principal quantum numbers, the optical periods associated with the generated atomic transitions are large (e.g., picoseconds for $n_i,n_f>10$) compared to the duration of the photoemitted electron pulses available in ultrafast electron microscopes~\cite{FES15,PLQ15}, and thus, electrons prepared as short pulses can be regarded as classical point particles in their interaction with Rydberg atoms, amenable to synchronization with high temporal precision relative to the transition optical period. In our study, we consider the atom to be placed at a well-defined spatial position (e.g., trapped in an optical lattice). 

These considerations foresay a high degree of selectivity over the resulting final atomic state by playing with the electron kinetic energy, the time of interaction, and the e-beam arrangement relative to the atom. For example, we have shown that trains of electrons can be used to maximize the transition probability for designated transition energies. In addition, the predicted strong electron--atom coupling can be controllable through the e-beam parameters.

As an appealing opportunity opened by the interaction between controlled e-beams and Rydberg atoms, we argue that quantum entanglement between the electron and the atom can be achieved by laterally splitting the electron wave function into different paths, following electron-wave manipulation methods developed in the context of electron holography~\cite{STP17}. Moreover, projection and postselection of the electron enable the generation of entangled atom pairs and, more generally, atomic ensembles. This approach to multiatom entanglement exceeds the capabilities of laser-based far-field optics and could potentially be employed to obtain complex on-demand few-body states in a selection of a given number of Rydberg atoms.

In brief, free electrons constitute an appealing tool to manipulate Rydberg atoms with capabilities that exceed those of far-field optical methods, such as the induction of nondipolar transitions, the synthesis of complex final atomic states, the control of such states through the energy and spatial profile of the free electrons, and the generation of entanglement between the electron and the atom as well as among two or more atoms traversed by the electron.

\appendix
\section{}
\renewcommand{\theequation}{S\arabic{equation}}
\label{appendixA}

We solve Eq.~(\ref{eq:SE-alphas}) by direct time integration using the nonretarded ($c\rightarrow\infty$) matrix elements
\begin{align}
M_{ij}(t) = e^2\int \mathrm{d}^3\rb\, \frac{\psi_i^*(\rb) \,\psi_j(\rb)}{|\rb-\rb_0(t)|} \label{Mij}
\end{align}
associated with $\ket{n_j,l_j,m_j} \rightarrow \ket{n_i,l_i,m_i}$ transitions produced by an electron following a trajectory given by $\rb=\rb_0(t)$. These elements admit analytical expressions when using real-space hydrogenic wave functions
\begin{align}
&\psi_{n,l,m}(\rb)=\braket{\rb|n,l,m} \nonumber\\
&= \sqrt{\bigg(\frac{2}{n a_0}\bigg)^3\frac{(n-l-1)!}{2n (n+l)!}} \ee^{-r/n a_0}\, \bigg(\frac{2r}{n a_0}\bigg)^l L_{n-l-1}^{2l+1}\bigg(\frac{2r}{n a_0}\bigg)\, Y_{l,m}(\Omega_{\rr}),
\nonumber
\end{align}
where $L_\alpha^\beta$ are generalized Laguerre polynomials and $Y_{l,m}$ are spherical harmonics satisfying the orthogonality relation $\int\mathrm{d}^2\Omega\, Y^*_{lm}(\Omega)Y_{l'm'}(\Omega)= \delta_{ll'}\delta_{mm'}$. More precisely, expanding the generalized Laguerre polynomials into powers of $r$, we can write $\psi_i^*(r)\psi_j(r)= \ee^{-g r}\sum_k A_k r^k$ using a finite number of coefficients $A_k$ indexed by integers $k$ and defining $g=(1/n_i+1/n_j)/a_0$. Inserting this expression into Eq.~(\ref{Mij}) together with the expansion $1/|\rb-\rb_0|=4\pi\sum_{l=0}^\infty\sum_{m=-l}^{l} (2l+1)^{-1}\,\big(r_<^l/r_>^{l+1}\big)\, Y_{lm}(\Omega_{\rb})Y^*_{lm}(\Omega_{\rb_0})$ for the Coulomb interaction, where $r_>=\max\{r,r_0\}$ and $r_<=\min\{r,r_0\}$, we finally obtain
\begin{align}
M_{ij}(t) =4\pi e^2\sum_{klm} \frac{A_k}{2l+1} \,B_{l_jm_j,l_im_i,lm} \,C_{kl}(g,r_0)\,Y^*_{lm}(\Omega_{\rb_0}), \nonumber
\end{align}
where the angular integral $B_{lm,l'm',l''m''}=\int {\rm d}^2\Omega\, Y^*_{lm}(\Omega) Y_{l'm'}(\Omega)Y_{l''m''}(\Omega)$ can be expressed in terms of Clebsch–Gordan coefficients~\cite{M1966}, while the radial integral yiels
\begin{align}
&C_{k,l}(g,r_0) = \int_0^\infty \mathrm{d}r \,\ee^{-g r}\,r^k \,\frac{r_<^l}{r_>^{l+1}} \nonumber\\
&=\frac{1}{g^k} \,\Big[(gr_0)^{-l-1}\,\gamma(k+l+1,g r_0) +(gr_0)^l\,\Gamma{(k-l,g r_0)}\Big] \nonumber
\end{align}
in terms of the incomplete and lower-incomplete gamma functions~\cite{AS1972} $\Gamma(\alpha,z)$ and $\gamma(\alpha,z)$.


%

\section*{Acknowledgements}
We are indebted to Claus Ropers and Murat Sivis for motivating the present study and suggesting that the interaction between free electrons and Rydberg atoms can produce large excitation probabilities. We thank them and Valerio Di Giulio for helpful and enjoyable discussions. This work has been supported in part by the European Research Council (Advanced Grants 789104-eNANO and 101055435-ULEEM), the European Commission (Horizon 2020 Grants No. 101017720 FET-Proactive EBEAM and No. 964591-SMART-electron), the Spanish MICINN (PID2020-112625GB-I00 and Severo Ochoa CEX2019-000910-S), the Catalan AGAUR (Grant No. 2024 FI-2 00052) and CERCA Programs, and Fundaci\'{o}s Cellex and Mir-Puig.

\pagebreak \onecolumngrid
\section*{Additional figures}
\renewcommand{\thefigure}{S\arabic{figure}}
\setcounter{figure}{0}

\begin{figure*}[htb] \label{FigS1}
\centering \includegraphics[width=0.9\textwidth]{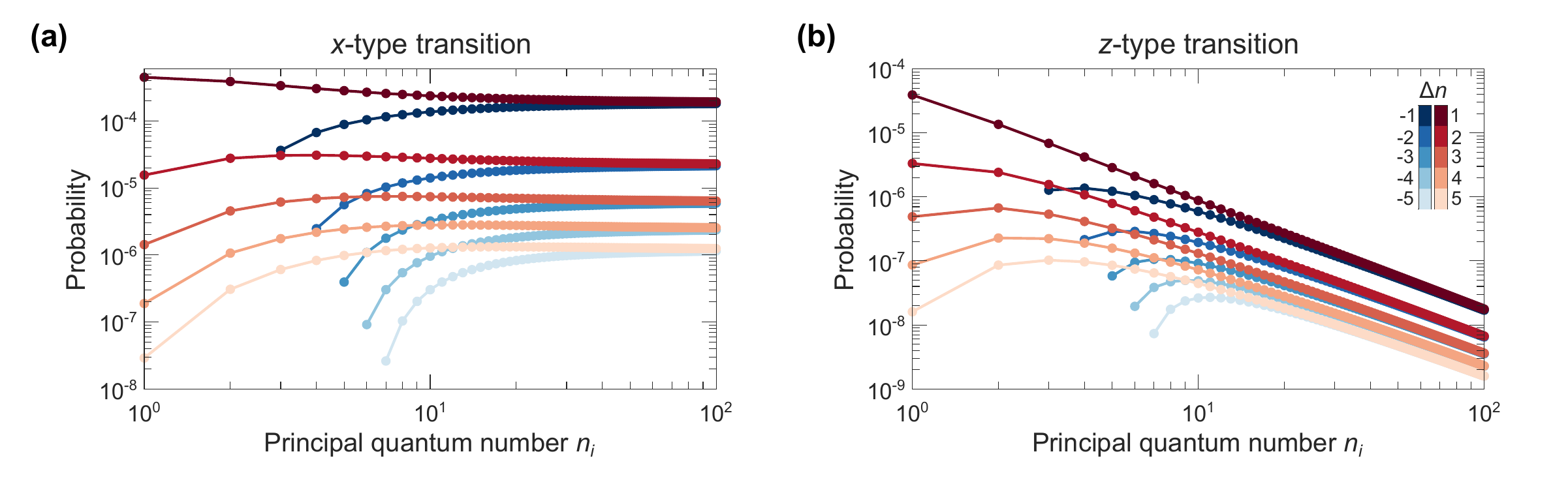}
\caption{{\bf Dependence of the dipolar transition probability on principal quantum numbers and the orientation of the transition dipole.} We plot the first-order dipolar transition probability produced by a free electron moving with velocity $v=0.1\,c$ (parallel to the quantization axis $z$) and impact parameter $b=\big(n_i^2 + n_f^2\big)\,a_0$ from an initial s state $\ket{n_i,0,0}$ to final p states [$(1/\sqrt{2})\big(\ket{n_i+\Delta n,1,1}+\ket{n_i+\Delta n,1,-1}\big)$ (a, $x$-type transition) and $\ket{n_i+\Delta n,1,0}$ (b, $z$-type transition)] for various values of $\Delta n$ [see legend in (a)].}
\end{figure*}

\begin{figure*}[htb] \label{FigS2}
\centering \includegraphics[width=0.8\textwidth]{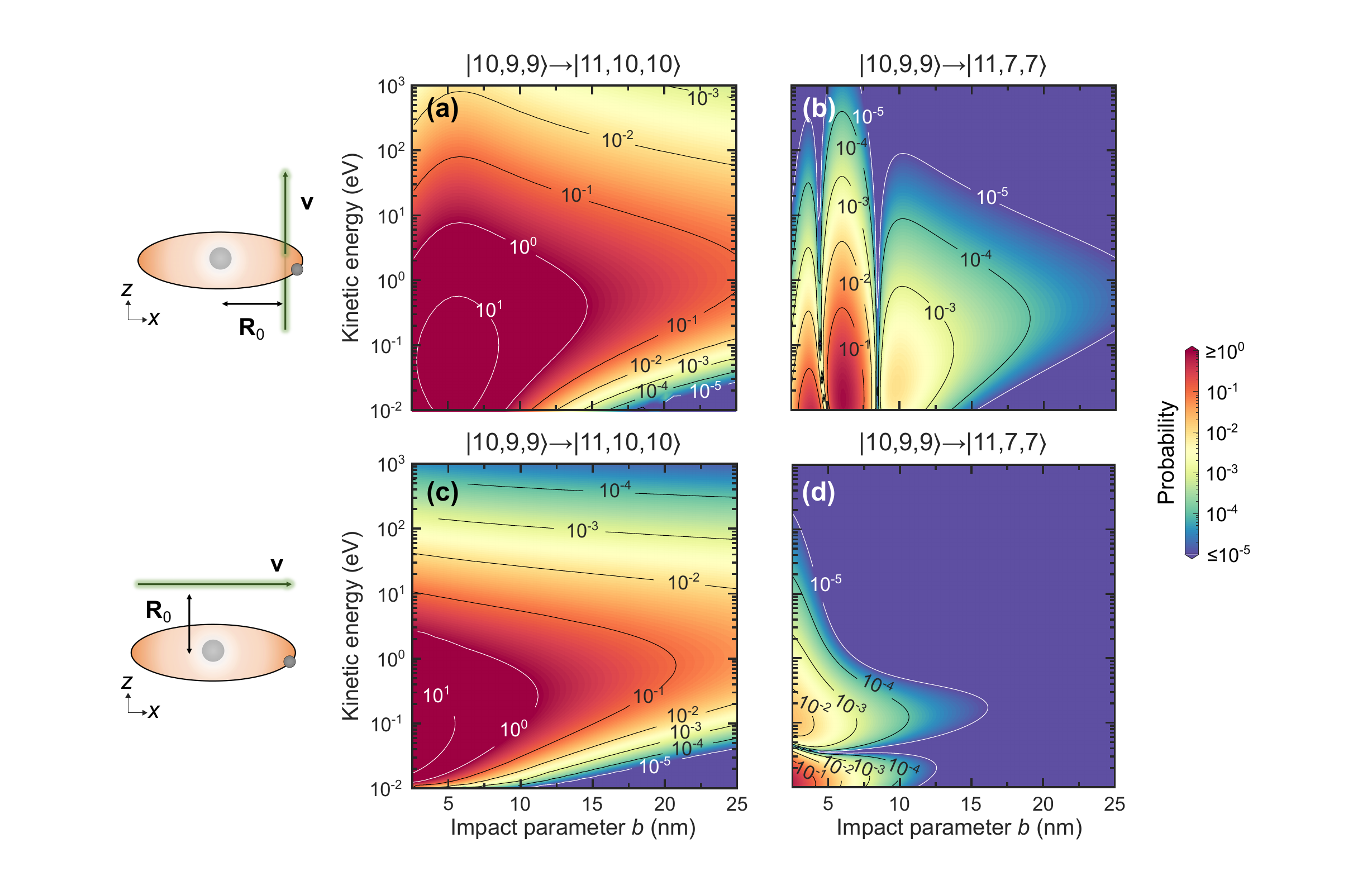}
\caption{\textbf{Strong coupling regime and nondipolar transitions revealed within first-order perturbation theory.} We plot the transition probabilities calculated within first-order perturbation theory for the e-beam configurations depicted in left insets: velocity and impact parameter $\vb = v\zz$ and $\Rb_0 = b\xx$ in (a,b), and $\vb = v\xx$ and $\Rb_0 = b\zz$ in (c,d). The quantization axis is $z$ in all cases. Starting in an initial state $\ket{10,9,9}$, the probability is represented as a function of the impact parameter $b$ and the electron kinetic energy for dipolar transitions (a,b; final state $\ket{11,10,10}$) and nondipolar transitions (a,b; final state $\ket{11,7,7}$).}
\end{figure*}

\begin{figure*}[htb] \label{FigS3}
\centering \includegraphics[width=0.95\textwidth]{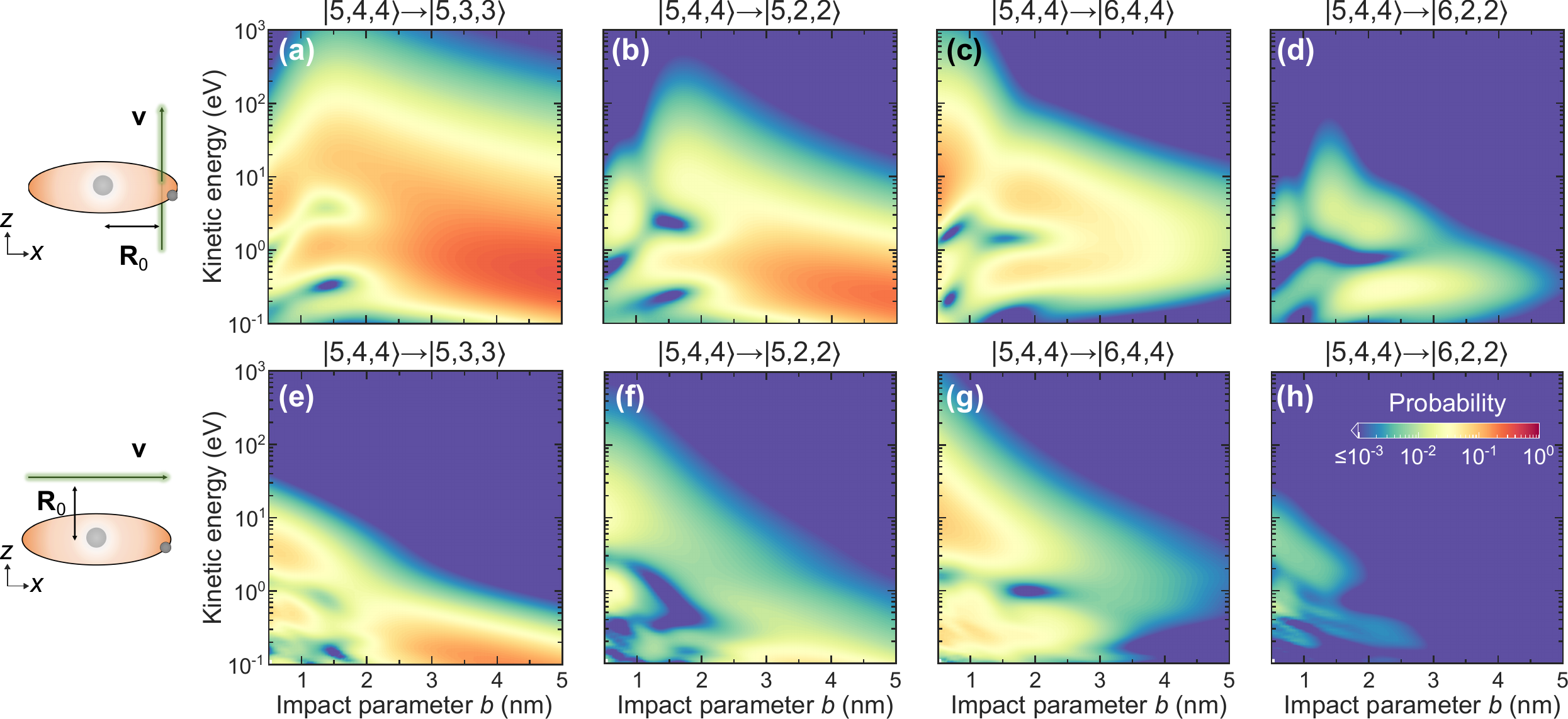}
\caption{\textbf{Additional transition probabilities under the conditions of Fig.~2 in the main text.} We plot probability maps for transitions from an initial state $\ket{5,4,4}$ to other final states calculated to all orders in the electron--atom interaction. We show results for
(a,b,e,f) elastic transitions to final states $\ket{5,3,3}$ and $\ket{5,2,2}$;
(c,g) inelastic transitions with no angular-momentum exchange (final state $\ket{6,4,4}$); and
(d,h) inelastic transitions with angular momentum exchanges ($\ket{6,2,2}$)
}
\end{figure*}

\end{document}